\documentclass{emulateapj}
\usepackage{graphicx} 
\usepackage{color}

\def\be{\begin{equation}} 
\def\ee{\end{equation}}

\def\HI{\hbox{H~$\scriptstyle\rm I\ $}} 
\def\HII{\hbox{H~$\scriptstyle\rm II\ $}}

\def\gsim{\lower.5ex\hbox{\gtsima}} 
\def\lsim{\lower.5ex\hbox{\ltsima}} \def\gtsima{$\; \buildrel > \over \sim \;$} \def\ltsima{$\; \buildrel < \over \sim \;$} \def\prosima{$\; 
\buildrel \propto \over \sim \;$} \def\gsim{\lower.5ex\hbox{\gtsima}} 
\def\lsim{\lower.5ex\hbox{\ltsima}} 
\def\simgt{\lower.5ex\hbox{\gtsima}} 
\def\simlt{\lower.5ex\hbox{\ltsima}} 
\def\simpr{\lower.5ex\hbox{\prosima}}   
  
 \def\gtsima{$\; \buildrel > \over \sim \;$} 
\def\ltsima{$\; \buildrel < \over \sim \;$} 
\def\gsim{\lower.5ex\hbox{\gtsima}} 
\def\lsim{\lower.5ex\hbox{\ltsima}} 
\def\simgt{\lower.5ex\hbox{\gtsima}} 
\def\simlt{\lower.5ex\hbox{\ltsima}} 
\def\simpr{\lower.5ex\hbox{\prosima}}

\def\E3{{\cal E}_{\rm g}^{III}}

\def\Msun{\rm M_\odot}
\def\Zsun{\rm Z_\odot}

\def\Msun{\rm M_\odot}
\def\Zsun{\rm Z_\odot}
\def\M*{M_*}
\def\Z*{Z_*}
\def\L*{L_*}
\def\MUV{M_{UV}}

\def\fesc{f_{esc}}
\def\kev{\,\rm{keV}} 
\def\mx{\,m_x} 
\def\highz{high-$z$\,}

\shorttitle{Reionization and  in galaxy formation in WDM}
\shortauthors{Dayal et al.}

\begin{document}

\title{Reionization and galaxy formation in Warm Dark Matter cosmologies}
\author{Pratika Dayal\altaffilmark{1}, Tirthankar Roy Choudhury\altaffilmark{2}, Volker Bromm\altaffilmark{3} \& Fabio Pacucci\altaffilmark{4}}
\altaffiltext{1}{Kapteyn Astronomical Institute, University of Groningen, P.O. Box 800, 9700 AV Groningen, The Netherlands}
\altaffiltext{2}{National Centre for Radio Astrophysics, Tata Institute of Fundamental Research, Pune 411007, India}
\altaffiltext{3}{Department of Astronomy and Texas Cosmology Centre, University of Texas, Austin, TX 78712, USA}
\altaffiltext{4}{Department of Physics, Yale University, P.O. Box 208121, New Haven, CT 06520, USA }

\begin{abstract}
{We compare model results from a semi-analytic (merger-tree based) framework for high-redshift ($z \simeq 5-20$) galaxy formation against reionization indicators, including the {\it Planck} electron scattering optical depth ($\tau_{es}$) and the ionizing photon emissivity ($\dot n_{ion}$), to shed light on the reionization history and sources in Cold (CDM) and Warm Dark Matter (WDM; particle masses of $\mx = 1.5,3$ and 5 keV) cosmologies. This model includes all the key processes of star formation, supernova feedback, the merger/accretion/ejection driven evolution of gas and stellar mass and the effect of the ultra-violet background (UVB), created during reionization, in photo-evaporating the gas content of galaxies in halos with $M_h \lsim 10^9 \Msun$. We find that the delay in the start of reionization in light (1.5 keV) WDM models can be compensated by a steeper redshift evolution of the ionizing photon escape fraction and a faster mass assembly, resulting in reionization ending at comparable redshifts ($z \simeq 5.5$) in all the Dark Matter models considered. We find the bulk of the reionization photons come from galaxies with a halo mass $M_h \lsim 10^9\Msun$ and a UV magnitude $ -15 \lsim M_{UV} \lsim -10$ in CDM. The progressive suppression of low-mass halos with decreasing $\mx$ leads to a shift in the ``reionization" population to larger halo masses of $M_h \gsim 10^9\Msun$ and $ -17 \lsim M_{UV} \lsim -13$ for 1.5 keV WDM. We find that current observations of $\tau_{es}$ and the Ultra-violet luminosity function are equally compatible with all the (cold and warm) Dark Matter models considered in this work. Quantifying the impact of the UVB on galaxy observables (luminosity functions, stellar mass densities, stellar to halo mass ratios) for different DM models we propose that global indicators including the redshift evolution of the stellar mass density and the stellar mass-halo mass relation, observable with the {\it James Webb Space Telescope}, can be used to distinguish between CDM and WDM (1.5 keV) cosmologies. }

\end{abstract}


\section{Introduction}
According to the standard Lambda Cold Dark Matter ($\Lambda$CDM) cosmological model the universe consists of three main components: Dark Energy, Cold Dark Matter (CDM) and baryons with energy densities corresponding to $\Omega_\Lambda = 0.6911$, $\Omega_m = 0.3089$ and $\Omega_b = 0.049$ respectively \citep{planck2015}. Dark Matter forms the basis of the hierarchical model of galaxy formation, providing the halos into which baryonic matter collapses, merging and accreting matter to form successively larger structures through cosmic time. With its property of clustering on all scales, CDM has been remarkably successful in predicting the large scale structure of the universe, the temperature anisotropies measured by the Cosmic Microwave Background (CMB) and Lyman-$\alpha$ forest statistics \citep[e.g.][]{peebles1971, blumenthal1984, bond-szalay1983, cole2005, lange2001, hinshaw2013, planck2014, slosar2013}. However, as recently reviewed by \citet{weinberg2013}, CDM exhibits a number of small scales problems including: (i) producing halo profiles that are cuspy as opposed to the observationally preferred constant density cores \citep{navarro1997, subramanian2000}, (ii) over-predicting the number of satellite and field galaxies as compared to observations - the ``missing satellite problem"\citep {klypin1999, moore1999}, (iii) predicting massive (Large Magellanic Cloud mass) concentrated Galactic sub-halos inconsistent with observations \citep[e.g.][]{boylan2012} and (iv) facing difficulties in producing typical disks due to ongoing mergers down to $z \simeq 1$ \citep{wyse2001}. The limited success of baryonic feedback in solving these small scale problems \citep[e.g.][]{boylan2012, teyssier2013} has prompted questions regarding the nature of dark matter itself. Indeed, a popular solution to the small scale problems in CDM cosmology involves invoking ($\sim$ keV) Warm Dark Matter (WDM) particles that erase small-scale power \citep[e.g]{blumenthal1984, bode2001}~\footnote{However, some works caution that the extremely low mass WDM particles required to make constant density cores prevent the very formation of  dwarf galaxies \citep{maccio2012_catch22, schneider2014}.}. Particle physics provides compelling motivation for WDM candidates such as sterile neutrinos which can be as light as $\sim {\cal O}(1)$~keV: extensions beyond the Standard Model allow $3$ sterile neutrinos, $2$ of which could be as heavy as $1-10$~GeV, with the lightest one being of the order of ${\cal O}(1)$ keV \citep[for a review see][]{Abazajian2012}. The possibility of sterile neutrinos being a viable dark matter candidate has been recently lent support by observations of a $3.5 \kev$ monochromatic line, that might arise from a light sterile neutrino annihilating into photons, from the Perseus galaxy cluster detected using XMM-Newton \citep{bulbul2014, boyarsky2014}.

The most stringent astrophysical constraints on the WDM particle mass of $\mx \geq 3.3 \kev$ have been obtained using the low-$z$ ($z \gsim 4$) Lyman-$\alpha$ flux power spectrum \citep{viel2013}. However, allowing for the intergalactic medium (IGM) temperature-density relation to vary independently as a function of redshift, \citet{garzilli2015} have obtained a lower bound of $\mx \geq 2.2 \kev$ using the same data. A number of studies have used alternative high-$z$ probes to obtain constraints on the WDM particle mass ranging from $\mx \gsim 1.6 \kev$ using number counts of high-$z$ gamma ray bursts \citep{desouza2013} to $\mx \gsim 1\kev$ using dwarf spheroidal galaxy observations \citep{Devega2010}, simultaneously reproducing stellar mass functions and the Tully-Fisher relation for $z=0-3.5$ galaxies \citep{kang2013} and comparing the observed number density of $z\approx10$ galaxies to that expected from the halo mass function \citep{pacucci2013}. 

With its property of smearing out small-scale density perturbations, WDM leads to a delay in the assembly of the lowest mass galaxies at early epochs, resulting in a corresponding delay in the appearance of the first metal-free Population III (PopIII) stars. In addition, since low-mass high-$z$ galaxies are believed to be the main sources of reionizing photons in the early universe \citep[e.g.][]{barkana-loeb2001, ciardi-ferrara2005, choudhury-ferrara2007, cfg2008, bouwens2015b, robertson2015}, their delayed and accelerated assembly would naturally lead to a corresponding delay and acceleration in the reionization history. A number of works have used such arguments to constrain the WDM particle mass: \citet{yoshida2003} have used simulations of PopIII star formation to show that they can yield a maximum electron scattering optical depth of $\tau_{es} \simeq 0.06$ in WDM models with masses as large as $\mx \simeq 10 \kev$. While such models could be ruled out using  
 WMAP data, this number is in good agreement with the latest {\it Planck} estimates \citep{planck2015, planck2016}. \citet{magg2016} have recently proposed observing the PopIII supernovae (SN) rate with the forthcoming {\it James Webb Space telescope} (JWST) and the {\it European-Extremely Large Telescope} (E-ELT) to differentiate between CDM and WDM, and constrain the mass of the latter. \citet{somerville2003} have used the global star formation rate density (SFRD) to show that reionization is delayed until $z \simeq 6$ in  $\mx \simeq 1.5 \kev$ models, a result that is broadly consistent with the latest {\it Planck} results. \citet{yue2012} have used a minimum halo mass for star formation embedded in a bubble-model for reionization to show that although WDM can delay the start of reionization, the end might not necessarily be delayed; however, these authors assume a fixed ionizing photon production rate per baryon and do not explicitly model any baryonic physics. Recently, using a constant relation linking the ultra-violet (UV) and ionizing photon production rates to the DM halo mass, \citet{schultz2014} have ruled out $\mx \lsim 1.3\kev$; however, even $\mx \simeq 0.8\kev$ WDM yields an optical depth consistent with the latest {\it Planck} results \citep{planck20142, planck2016} using their model. Finally, using a semi-analytic model of galaxy formation where 10\% of the baryons that collapsed into halos, before virialization of the outermost DM shell, are converted into stars \citet{barkana2001} obtain constraints of $\mx \gsim 1.2 \, (0.4) \kev$ if the star formation efficiency is the same (10 times larger) at high-$z$ as at $z \sim 3$. 
 
 This last work is most similar in spirit to our approach, although it does not include gas loss from SN winds which progressively decreases the gas fraction in galaxies as shown in \citet{dayal2014}.  A major caveat in many of these existing works is their assumption of a cosmology-independent mass-to-light (M/L) or ionizing photon production rate-to-mass ratio. However, as shown in \citet{dayal2014_wdm1}, the M/L ratio sensitively depends on the cosmology considered: a delay in structure formation in WDM models results in a younger stellar population in low-mass halos that, as a result, produces more light per unit stellar mass compared to CDM. The same behaviour is expected to hold for the amount of reionization photons produced per unit stellar mass at any cosmic epoch.

Over the past few years, ground and space-based observations have allowed a statistically significant data-set to be collected for $z \simeq 6-10$ Lyman Break galaxies \citep[LBGs;][]{oesch2010, bouwens2010a,bouwens2011b, castellano2010, mclure2010, bradley2012, oesch2013, mclure2013, bowler2014a, bouwens2014, bowler2014b, atek2015, livermore2016, oesch2016}. In this work, our aim is to revisit the constraints allowed on the WDM particle mass by combining these ``galaxy" data sets with the latest limits on the cosmological electron scattering optical depth inferred by {\it Planck}. We build upon a semi-analytic model that traces both the DM and baryonic assembly of \highz ($\simeq 5-15$) galaxies, reproducing a number of key observables, presented in \citet{dayal2014, dayal2014_wdm1} to show how the reionization history and sources depend on the DM model considered. The strength of the model lies in the fact that in addition to internal feedback from SN \citep[see Sec. 2.3][]{dayal2014_wdm1} our model also includes the effects of ``external'' feedback (i.e. reionization photo-evaporating gas from DM halos).

The cosmological parameters used in this work correspond to ($\Omega_{\rm m },\Omega_{\Lambda}, \Omega_{\rm b}, h, n_s, \sigma_8) = (0.3089,0.6911,0.049, 0.67, 0.96, 0.81)$, consistent with the latest results from the {\it Planck} collaboration \citep{planck2015} and we quote all quantities in comoving units unless stated otherwise.

\section{Modelling galaxy formation and reionization}
In this work, we explore 4 DM models: CDM and WDM with $\mx =1.5,3$ and 5\kev. Although we cite $\mx$ values assuming thermally decoupled relativistic particles, these numbers can be converted into sterile neutrino masses ($m_{sterile\, \nu}$) using the expression \citep{viel2005}:
\begin{equation}
m_{sterile\, \nu} = 4.43 \kev \bigg(\frac{\it m_x}{1 \kev}\bigg)^{4/3} \bigg(\frac{0.1225}{\Omega_m {\it h}^2}\bigg).
\end{equation}
This yields $m_{sterile\, \nu} = (6.7,16.8,33.3)\kev$ corresponding to $\mx = (1.5,3,5)\kev$ respectively. 

\subsection{Merger trees and the baryonic implementation}
We now briefly summarise the theoretical model and interested readers are referred to \citet{dayal2014, dayal2014_wdm1} for complete details. We construct 400 merger trees starting at $z=4$, linearly distributed across the halo mass range $\log(M_h/\Msun) = 9-13$ for the 4 DM models considered. The merger-trees use 320 equal redshift steps ($\Delta z = 0.05$) between $z=20$ and $z=4$ with a mass resolution of $M_{res} =10^8 \Msun$ using the modified binary merger tree algorithm with smooth accretion detailed in \citet{parkinson2008} and \citet{benson2013}. We scale the relative abundances of the merger tree roots to match the $z=4$ Sheth-Tormen halo mass function \citep[HMFs;][]{sheth-tormen1999} and have verified that these yield HMFs in good agreement with the Sheth-Tormen HMF at all $z$. 

We implement the merger trees with baryonic physics including star formation, SN feedback, and the merger/accretion/ejection driven evolution of gas and stellar masses. Our model is based on the simple premise that any galaxy can form stars with a maximum efficiency ($f_*^{ej}$) that provides enough energy to expel all the remaining gas,  quenching further star formation, up to a threshold value of $f_*$ \citep[see][]{dayal2014}. This implies the {\it effective} star formation efficiency ($f_*^{eff}$) is the minimum between $f_*^{ej}$ and $f_*$. This model has two $z$- and mass-independent free parameters whose values are selected to match the evolving ultra-violet luminosity function (UV LF): the maximum threshold star formation efficiency ($f_*$) and the fraction of SN energy that goes into unbinding gas ($f_w$). We implement this simple idea proceeding forward in time from the highest merger tree output redshift, $z=20$. At any $z$ step, the gas mass in a galaxy is determined both by the gas mass brought in by merging progenitors as well as that smoothly-accreted from the IGM. A part ($f_*^{eff}$) of this gas forms new stellar mass, $M_*(z)$, with the final gas mass depending on the ratio of the (instantaneous) energy provided by exploding SN and the potential energy of the halo. Further, at any step the total stellar mass in a galaxy is the sum of newly-formed stellar mass, and that brought in by its progenitors. In this work we also explore the effects of the ultra-violet background (UVB) in photo-evaporating gas from low-mass halos, impeding their star-formation capabilities, and its impact on both galaxy assembly and reionization as explained in Sec. \ref{sec_reio} that follows. 

For simplicity, we assume every new stellar population to have a fixed metallicity of $0.05 \Zsun$ and an age $t_0=2\, {\rm Myr}$. Using the population synthesis code {\small STARBURST99} \citep{leitherer1999}, the initial UV luminosity (at $\lambda = 1500$\,\AA) can be calculated as $L_{UV}(0) = 10^{33.077} (\M*/\Msun)\, {\rm erg \, s^{-1} \AA^{-1}}$ and the initial output of ionizing photons can be calculated as $\dot n_{int}(0) = 10^{46.6255} (M_*/\Msun) s^{-1}$. Further, the time evolution of these quantities can be expressed as
\begin{equation}
L_{UV}(t) = L_{UV}(0) -1.33 \log\frac{t}{t_0} + 0.462 
\end{equation}

\begin{equation}
\dot n_{int}(t) = \dot n_{int}(0) -3.92 \log \frac{t}{t_0} + 0.7. 
\label{uv_nion}
\end{equation}
For any galaxy along the merger tree the UV luminosity and ionizing photon output rate are the sum of the values from the new starburst and the contribution from older populations accounting for the drop with time. 

As shown in \citet{dayal2014, dayal2014_wdm1}, our model reproduces the observed UV LF for all DM models (CDM and WDM with $\mx=1.5,3$ and 5\kev) at $z \simeq 5-10$ over 7 magnitudes in luminosity and predicts the $z$-evolution of the faint end UV LF slope, in addition to reproducing key observables including the stellar mass density (SMD) and mass-to-light ratios using fiducial parameter values of $f_* = 0.038$ and $f_w = 0.1$. We maintain these fiducial parameter values in all the calculations carried out in this work. 

\begin{figure*}
\center{\includegraphics[scale=0.86]{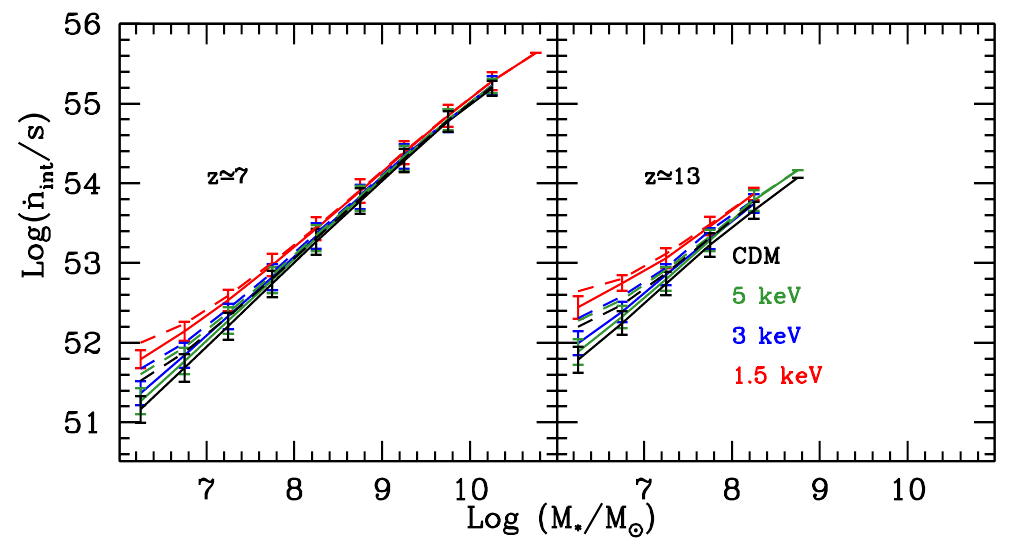}}
\caption{Intrinsic production rate of hydrogen ionizing photons as a function of the stellar mass at $ z \simeq 7$ and 13 in the left and right panels, respectively. In each panel the solid lines show results considering SN feedback only; dashed lines show results considering both SN and UV feedback in the fiducial model ($M_{gas}=0$ for $M_h \lsim 10^9 \Msun$ halos). The $1-\sigma$ error bars for each $M_*$ bin are plotted for the SN-feedback only scenario. While $\dot n_{int}$ scales with the stellar mass in all DM models, 1.5 keV galaxies produce more ionizing photons per unit stellar mass as a result of their younger stellar populations. Including the effects of UV feedback suppresses star formation in low-mass halos resulting in a lower final stellar mass; however, the final $\dot n_{int}$ value, that depends on the latest burst of stars formed, remains unaffected (see Sec. \ref{nion} for details).}
\label{nion_fnms} 
\end{figure*}

\subsection{Modelling reionization}
\label{sec_reio}
The reionization history, expressed through the evolution of the volume filling fraction ($Q_{II}$) for ionized hydrogen (\HII), can be written as \citep{shapiro-giroux1987, madau1999}
\begin{equation}
\label{filfracz}
\frac{dQ_{II}}{dz} = \frac{dn_{ion}} {dz}\frac{1}{n_H} - \frac{Q_{II}}{t_{rec}} \frac{dt}{dz},
\end{equation}
where the first term on the right hand side represents the growth of \HII regions while the second term accounts for the decrease in $Q_{II}$ due to recombinations. Here, $dn_{ion}/dz = f_{esc} dn_{int}/dz$ represents the hydrogen ionizing photon rate density (per comoving volume) contributing to reionization, with $f_{esc}$ accounting for the fraction of ionizing photons that escape out of the galactic environment. Further, $n_H$ is the comoving hydrogen number density, $dt/dz = [H(z) (1+z)]^{-1}$ and $t_{rec}$ is the recombination time that can be expressed as \citep[e.g.][]{madau1999}
\begin{equation}
t_{rec} = \frac{1}{\chi\, n_H \, (1+z)^3 \alpha_B \, C}.
\end{equation}
Here $\alpha_B$ is the hydrogen case-B recombination coefficient, $\chi = 1.08$ accounts for the excess free electrons arising from singly ionized helium and $C$ is the IGM clumping factor. We use the results of \citet{pawlik2009} and \citet{haardt-madau2012} who show that the UVB generated by reionization can act as an effective pressure term, reducing the clumping factor with $z$ such that
\begin{equation}
C = \frac{<n_{HII}^2>}{<n_{HII}>^2} = 1+ 43 \, z^{-1.71}.
\end{equation}

\begin{table*}
\begin{center}
\caption{For each of the DM models considered (column 1) we show the redshift-evolution of the intrinsic production rate of hydrogen ionizing photons (column 5) for different UV magnitude limits (column 4) in the presence of both SN and UVB feedback (columns 2 and 3).}
\begin{tabular}{|c|c|c|c|c|}
\hline
DM Model & SN feedback & UVB feedback & UV cut & $\dot n_{int}-z$ relation\\
\hline
CDM & yes & yes &  all galaxies (Eqn. \ref{eqn_nint})& $\log (\dot n_{int}) = -0.0036 z^2 - 0.058 z+52$\\
3 keV WDM & yes & yes &  all galaxies (Eqn. \ref{eqn_nint})& $\log (\dot n_{int}) = -0.0088 z^2 - 0.037 z+52$\\
1.5 keV WDM & yes & yes &  all galaxies (Eqn. \ref{eqn_nint})& $\log (\dot n_{int}) = -0.026 z^2 - 0.13 z+52$\\
CDM & yes & yes &  $M_{UV} \lsim -17.7$ & $\log (\dot n_{int}) = -0.0036 z^2 - 0.058 z+52$\\
3 keV WDM & yes & yes &  $M_{UV} \lsim -17.7$ & $\log (\dot n_{int}) = -0.021 z^2 - 0.004 z+52$\\
1.5 keV WDM & yes & yes &  $M_{UV} \lsim -17.7$ & $\log (\dot n_{int}) = -0.033 z^2 +0.17 z+52$\\
\hline
\end{tabular}
\label{table1}
\end{center}
\end{table*}

While reionization is driven by the hydrogen ionizing photons produced by early galaxies, the UVB built up during reionization suppresses the baryonic content of galaxies by photo-heating/evaporating gas at their outskirts \citep{klypin1999, moore1999, somerville2002}, suppressing further star formation and slowing down the reionization process. In order to account for the effect of UVB feedback on $\dot n_{ion}$, in the fiducial model, we assume total photo-evaporation of gas from halos below $M_{min}=10^9 \Msun$ embedded in ionized regions at any $z$; we vary this limit between $M_{min}= 10^{8.5-9.5}\Msun$ to check the robustness of our results. In this ``maximal external feedback" scenario, halos below $M_{min}$ in ionized regions neither form stars nor contribute any gas in mergers. The globally averaged $\dot n_{ion}$ can then be expressed as 
\begin{equation}
\label{eq_nion}
  \dot n_{ion}(z) = f_{esc} [Q_{II}(z) \dot n_{II}(z) + [1-Q_{II}(z)] \dot n_{I}(z)],
\end{equation}
where $\dot n_{II}$ and $\dot n_{I}$ account for the intrinsic hydrogen ionizing photon production rate density within ionized and neutral regions respectively. While $\dot n_{I}$ contains contribution from all sources, $\dot n_{II}$ represents the case where sources below $M_{min}$ do not contribute to the ionizing photon budget. At the beginning of the reionization process the volume filled by ionized hydrogen is very small ($Q_{II}<<1$) and most galaxies are not affected by the UVB, so that $\dot n_{ion}(z) \approx f_{esc} \, \dot n_{I}(z)$. As $Q_{II}$ increases and reaches a value $\simeq 1$, all galaxies less massive than $M_{min}$ are feedback-suppressed so that $\dot n_{ion}(z) \approx f_{esc} \, \dot n_{II}(z)$.

Given that $\dot n_{ion}(z)$ is an output of the model, with $t_{rec}$ calculated as a function of $z$, $f_{esc}$ is the only free parameter in our reionization calculations. In this work $f_{esc}$ is assumed to be independent of the halo mass. As shown in Sec. \ref{results} that follows, simultaneously fitting to observations of the CMB electron scattering optical depth ($\tau_{es}$) and the ionizing emissivity require a $z$-dependent $f_{esc}$.

\section{Reionization in different DM cosmologies}
\label{results}
We now discuss model constraints on $\mx$ using observed values of the CMB electron scattering optical depth ($\tau_{es}$) and ionizing emissivity ($\dot n_{ion}$). We then present the corresponding reionization history and sources for both CDM and WDM, as described in what follows.

\subsection{Ionizing photon production efficiency in different DM cosmologies}
\label{nion}

The delay in structure formation in WDM models results in a much younger stellar population hosted by low-mass halos at any $z$, when compared to CDM. This naturally results in low-mass WDM halos producing more ionizing photons, both considering and ignoring the effects of the UVB. This is the behaviour shown by $\dot n_{int}$ as a function of stellar mass, $M_*$, in Fig. \ref{nion_fnms}. We start by focusing on the left panel corresponding to $z \simeq 7$ where the $\dot n_{int}-M_*$ relation is essentially independent of the underlying DM model for galaxies with $M_* \gsim 10^9 \Msun$, corresponding to currently observed LBGs. However, given their younger ages, galaxies with $M_* \simeq 10^{6.2} \Msun$ produce about 4 times more ionizing photons in the 1.5 keV WDM model as compared to CDM. Quantitatively we find that $\log (\dot n_{int}) = 0.88 \log(M_*) + 46$ for the 1.5 keV scenario, steepening to $\log (\dot n_{int}) = \log(M_*) + 45$ for CDM, as also quantified in Table \ref{table1}; while the exact coefficients change, the same trends are seen at $z \simeq 13$ (right panel of the same figure).

\begin{figure*}
\center{\includegraphics[scale=0.86]{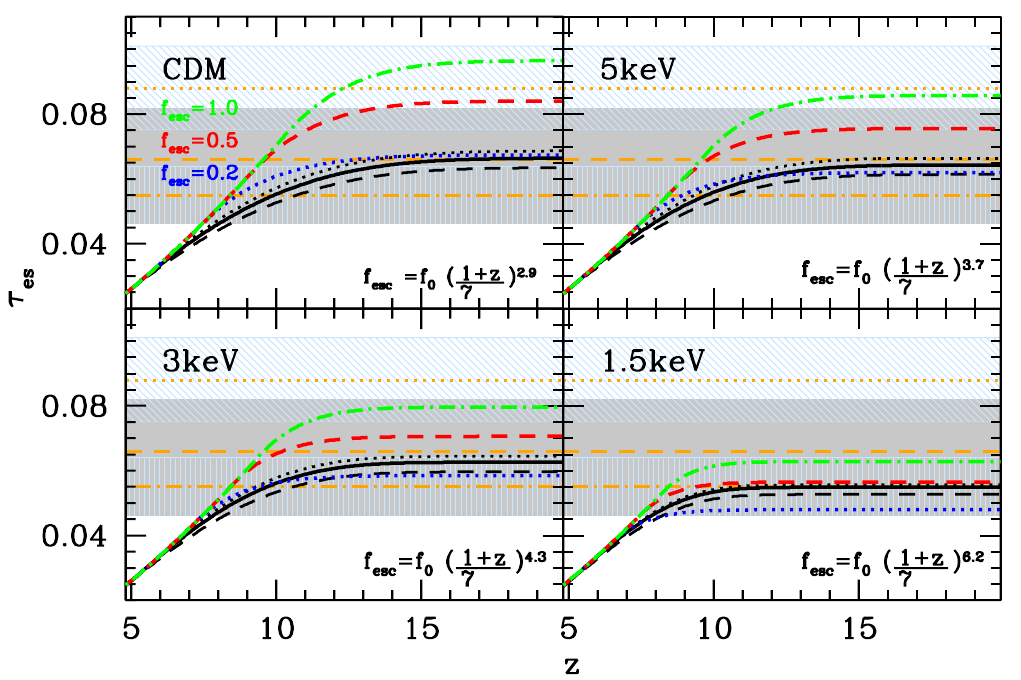}}
\caption{The CMB electron scattering optical depth ($\tau_{es}$) as a function of redshift for the 4 DM models considered in this paper, as marked. In each panel, lines show results using four different values of $f_{esc} =$ 1.0 (dot-dashed green line), 0.5 (dashed red line), 0.2 (dotted blue line) and the fiducial $z$-dependent value marked (solid black line). The dotted and dashed black lines show results using the fiducial $z$-dependent $f_{esc}$ value marked, but for a minimum reionization suppressed halo masses of $M_{min} = 10^{8.5} \Msun$ and $M_{min} = 10^{9.5} \Msun$, respectively; these show that varying $M_{min}$ within one order of magnitude does not change our results appreciably. The $z$-dependent $f_{esc}$ value has been obtained by simultaneously fitting to $\tau_{es}$ and ionizing photon emissivity observations (see Sec. \ref{joint_cons}). The horizontal dashed line shows the central value for $\tau_{es}$ inferred by Planck combining polarisation, temperature and lensing data \citet{planck2015} with the gray shaded region showing the $1-\sigma$ errors. The horizontal dotted and dot-dashed orange lines show the central value for $\tau_{es}$ inferred by {\it Planck} (2014) and {\it Planck} (2016), respectively; the inclined and vertically shaded (blue) regions show the associated $1-\sigma$ error bars. As seen, while the {\it Planck}-2014 results ruled out WDM as light as 1.5 keV, although requiring a very steep $z$-evolution of $f_{esc}$, this WDM model is allowed by the latest {\it Planck} (2015, 2016) results.  }
\label{fig_tau}
\end{figure*}

According to the fiducial model, including the effects of the UVB results in a loss of all gas content, switching-off star formation in all halos with $M_h \lsim 10^9 \Msun$. This naturally results in a lower stellar mass for the successors of such systems. On the other hand, $\dot n_{int}$ effectively only depends on the ongoing star formation rate (i.e. the instantaneous gas mass) given its steep decay with time as shown in Eqn. \ref{uv_nion}. As noted in \citet{dayal2014}, galaxies with $M_h \lsim 10^9 \Msun$ are feedback limited, contributing only stellar mass but no gas mass to their successors. So although the final stellar mass decreases in the presence of the UVB, $\dot n_{int}$ remains unchanged, resulting in a higher $\dot n_{int}-M_*$ relation as seen from the same figure. With the largest number of low-mass halos, the effect of UVB suppression is most strongly seen in CDM where $M_* \simeq 10^{6.2} \Msun$ halos produce about 2-2.5 times more ionizing photons as compared to the case with SN-feedback only; the effect of the UVB is the weakest  in the 1.5 keV case where $\dot n_{int}$ only increases by a factor of 1.5 for low-mass galaxies. Quantitatively, the $\dot n_{int}-M_*$ relation including the UVB is well-fit by $\log (\dot n_{int}) = 0.95 \log(M_*) + 46$ for CDM, changing to $\log (\dot n_{int}) = 0.85 \log(M_*) + 47$ for the 1.5 keV scenario. Again, galaxies at $z \simeq13$ follow the same qualitative trends.

\subsection{Joint reionization constraints from the CMB optical depth and UVB emissivity}
\label{joint_cons}
Combining {\it Planck} polarisation measurements with temperature and lensing data, the electron scattering optical depth corresponds to $\tau_{es} = 0.066 \pm 0.016$ \citep{planck2015}, compared to the value of $\tau_{es} = 0.089^{+0.012}_{-0.014}$ obtained by combining {\it Planck} and {\it WMAP} low-{\it l} polarization data \citep{planck20142}; low multipole EE data has been used to infer an even lower value of $\tau_{es} = 0.055 \pm 0.009$ \citep{planck2016}. We calculate $\tau_{es}$ as a function of $z$ by solving the equation:
\begin{equation}
\tau_{es}(z) = \sigma_T c \int_0^z dt \, n_e (1+z)^3
\label{tau}
\end{equation}
where $n_e (z) = Q_{II}(z) n_H$ is the global average comoving value of the electron number density and $\sigma_T = 6.6524 \times 10^{-25} {\rm cm}^2$ is the Thomson scattering cross-section. $f_{esc}(z)$ governs the evolution of $Q_{II}$ as shown by Eqns. \ref{filfracz} to \ref{eq_nion}. 

\begin{figure*}
\center{\includegraphics[scale=0.86]{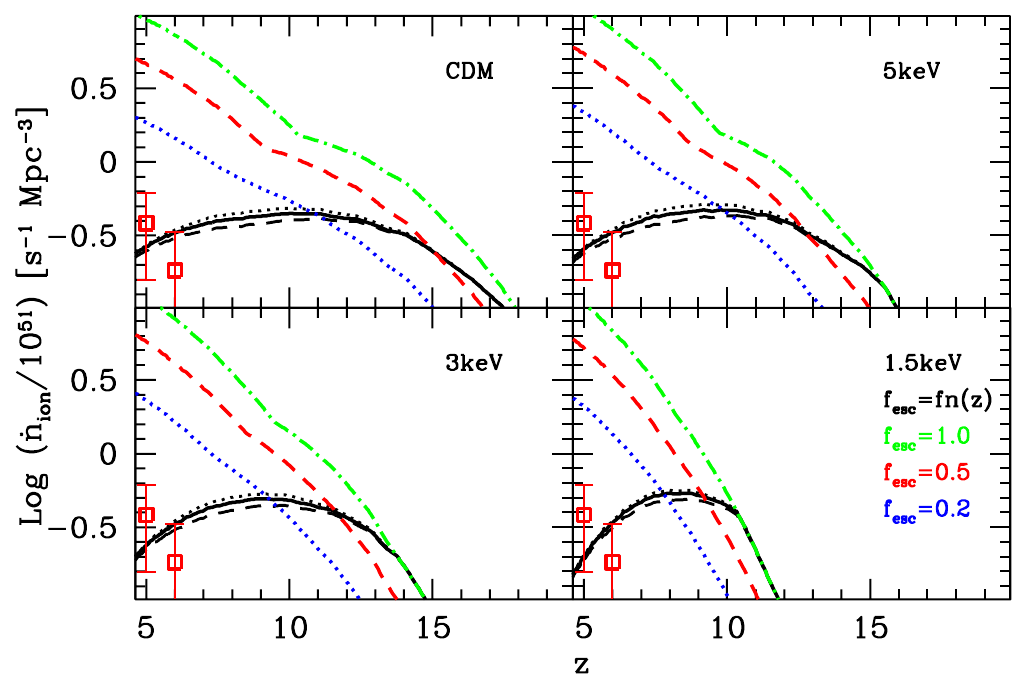}}
\caption{Redshift evolution of the \HI ionizing photon emissivity for the 4 DM models considered in this paper, as marked. In each panel, lines show results using four different values of $f_{esc} =$ 1.0 (dot-dashed green line), 0.5 (dashed red line), 0.2 (dotted blue line) and the fiducial $z$-dependent value (solid line). The dotted and dashed black lines show results using the fiducial $z$-dependent value marked, but for $M_{min} = 10^{8.5} \Msun$ and $M_{min} = 10^{9.5} \Msun$, respectively, to show that varying $M_{min}$ by one order of magnitude does not change our results appreciably. As seen, a constant $f_{esc}$ value severely over-predicts $\dot n_{ion}$ compared to the observations (points) at $z \simeq 5,6$ in all DM models. The observational results (and associated error bars) have been calculated following the approach of \citet{kfg2012}, i.e., by combining the observational constraints on $\Gamma_{HI}$ from \citet{wyithe-bolton2011} with the mean free path for ionizing photons ($\lambda_{mfp}$) from \citet{songaila-cowie2010}. See Sec. \ref{joint_cons} for details.}
\label{fig_q} 
\end{figure*}

A second observable that needs to be fit simultaneously is the emissivity of ionizing photons, which has been measured for $z \lesssim 6$. In this work, we follow the approach of \citet{kfg2012} and combine the observational constraints on the \HI photoionization rate ($\Gamma_{HI}$) from \citet{wyithe-bolton2011} with the mean free path of ionizing radiation ($\lambda_{mfp}$) from \citet{songaila-cowie2010} to obtain the observational estimate of $\dot n_{ion}$. We have used the fiducial values from \citet{kfg2012}: $\gamma=1$ for the source spectral index and $\beta=1.3$ for the \HI column density distribution. Varying these within allowed would only result in larger error bars, leaving our results unchanged \citep[see][]{kfg2012}. 

Once the parameters related to the galaxy formation model are fixed our model contains only one additional free parameter, $f_{esc}$, that can be varied to fit the two observational constraints mentioned above. Unfortunately there do not exist observations of $f_{esc}$ at high redshifts which can be used as possible prior bounds. The indirect constraints on $f_{esc}$ \citep[e.g.][]{mitra2013,robertson2015} have been obtained only for the standard CDM model and may not be valid for WDM cosmologies. We thus conservatively allow $f_{esc}$ to take any value between 0 and 1 and constrain it by simultaneously fitting $\tau_{es}$ and $\dot n_{ion}$ for each DM model. 

We start by assuming $f_{esc}$ to have a redshift-independent constant value for all the DM models considered \citep[e.g.][]{schultz2014}. As shown in Fig. \ref{fig_tau}, the {\it Planck} 2014 value of $\tau_{es} = 0.089^{+0.012}_{-0.014}$ \citep{planck20142} requires $f_{esc} \gsim 0.5$ for the CDM and 5 keV models and $f_{esc} \sim 1$ for the 3 keV scenario; the 1.5 keV model can be ruled out using this data since it results in an optical depth lower than these {\it Planck} limits even in the maximal case of $f_{esc}=1$. The lower {\it Planck-2015} value of $\tau_{es} = 0.066 \pm 0.016$ is well-reproduced for CDM and $\mx \lsim 3\kev$ WDM models for $f_{esc} \simeq 0.2$; the 1.5 kev case requires a higher value of $f_{esc} \gsim 0.5$ to be consistent with the observed bounds. We note that the most recent, and lowest, value of $\tau_{es} = 0.055\pm 0.009$ \citep{planck2016} is matched by all DM models for $f_{esc}$ values as low as $20\%$. 

Although matching the observed $\tau_{es}$, even the minimum $z$-independent constant value of $f_{esc}=0.2$ over-predicts the observed value of $\dot n_{ion}$ by about 0.8 dex at $z\simeq 6$ as shown in Fig. \ref{fig_q}. A scenario wherein $f_{esc}$ is $z$-independent is therefore ruled out by the emissivity constraints, irrespective of the DM model considered. In what follows, we show results using $f_{esc}$ calibrated to the {\it Planck} 2015 results \citep{planck2015} that has been obtained combining polarisation, lensing and temperature data; however these results are, within the error bars and hence, also valid for the latest {\it Planck} 2016 estimates as shown in Fig. \ref{fig_tau}.


Reconciling the optical depth and emissivity data sets requires a {\it $z$-dependent} value of $f_{esc}$ such that
\begin{equation}
f_{esc}(z) = {\rm min}\left[1, ~f_0 \bigg(\frac{1+z}{7}\bigg)^\alpha\right]~~~(z \geq 5),
\end{equation}
where $f_0$ and $\alpha$ are $z$-independent parameters, whose values are determined individually for each DM model considered, the results of which are shown in Table \ref{table2}. This formalisation requires $\alpha > 0$, which provides enough photons at high-$z$ to obtain the right value of $\tau_{es}$, whilst yielding reasonably low $\dot n_{ion}$ values at later times; this point has already been noted in previous works \citep[see e.g.,][]{mitra2011,mitra2012}. With the largest number of low mass halos available to provide \HI ionizing photons, CDM requires the most shallow $z$-dependence on $f_{esc}$. A decrease in the number of low mass halos with decreasing $\mx$ requires successively larger $f_{esc}$ values at early times, resulting in a steeper $z$-dependence. Finally, varying the minimum halo mass below which the UVB photo-evaporates the entire gas content by an order of magnitude between $10^{8.5}$ to $10^{9.5} \Msun$ has a minimal effect on both $\tau_{es}$ and $\dot n_{ion}$: as seen from Fig. \ref{fig_tau}, while $\tau_{es}$ changes by 0.005, $\dot n_{ion}$ changes by less than 0.1 dex. For this reason, we only show results for the fiducial value of $M_{min} = 10^9 \Msun$ in what follows. 

We end by noting that while the older {\it Planck} (2014) results could have ruled out the 1.5 keV scenario all models are equally probable given the latest CMB results \citep{planck2015, planck2016}, although the 1.5 keV model requires an exceptionally steep $z$-evolution of $f_{esc}$. 

\begin{table}
\begin{center}
\caption{The parameter values for the $z$-evolution of the escape fraction ($f_{esc}$) for different DM models using {\it Planck} 2015 data that combines polarisation, lensing and temperature data; the numbers in brackets show the sterile neutrino mass corresponding to $\mx$. The $z$-dependence of $f_{esc}$ is taken to be $f_{esc}(z) = f_0 [(1+z)/7]^{\alpha}$.}
\begin{tabular}{|c|c|c|}
\hline
Model & $f_0 \times 100$ & $\alpha$ \\
\hline
CDM & 4.5 & 2.9\\
$\mx = 5$ keV \, (33.3 keV)& 4.1 & 3.7\\
$\mx = 3$ keV \, (16.8 keV)& 3.8 & 4.3\\
$\mx = 1.5$ keV\, (6.7 keV) & 4.8 & 6.2\\
\hline
\end{tabular}
\label{table2}
\end{center}
\end{table}

\begin{figure}
\center{\includegraphics[scale=0.47]{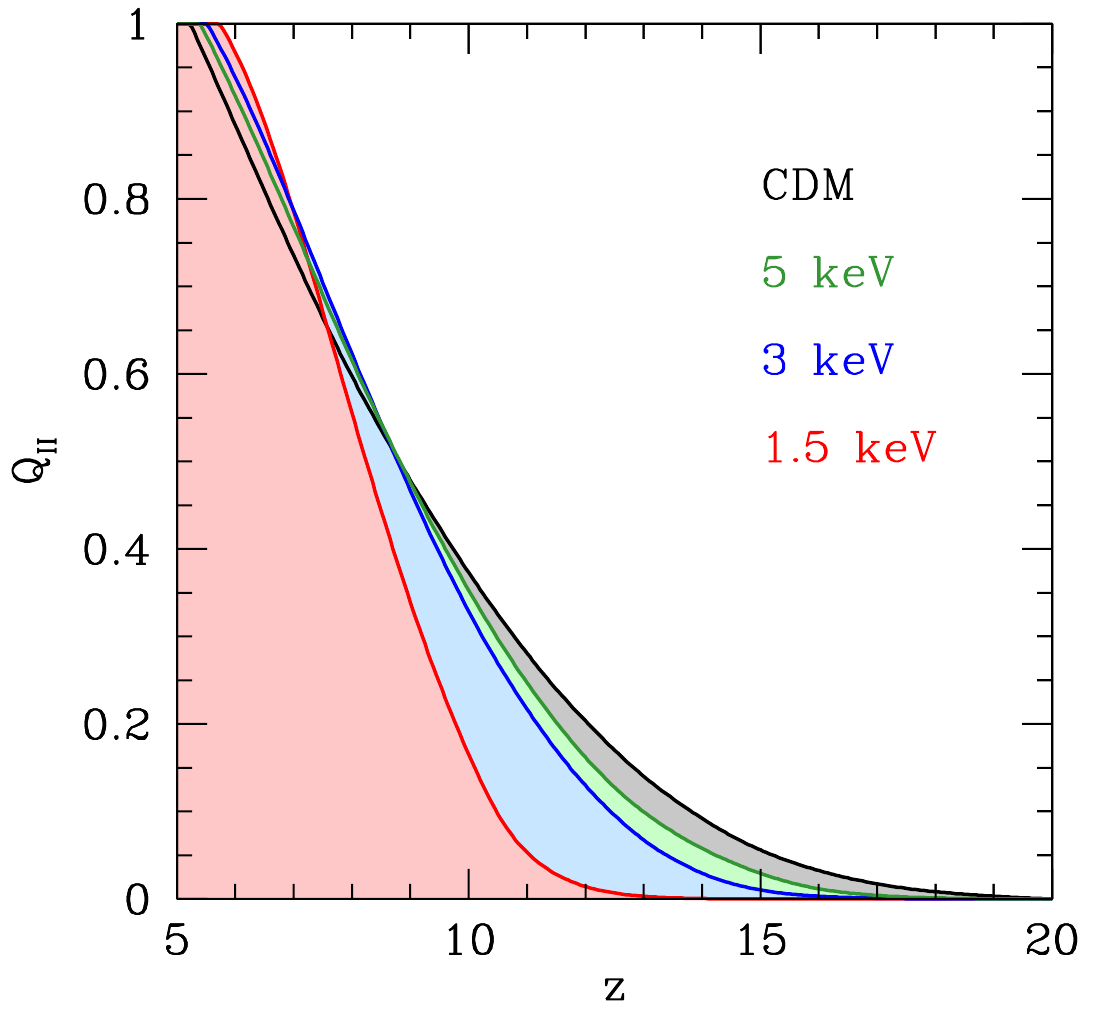}}
\caption{The volume filling fraction of ionized hydrogen as a function of $z$ for the fiducial parameter values for all 4 DM models considered in this work. While reionization starts at comparable epochs in CDM and $\mx \gsim 3 \kev$, the suppression of small-scale structure leads to a delay in reionization, by about 150 Myrs, for 1.5 keV WDM. However, a combination of accelerated galaxy assembly and a steeper $f_{esc}-z$ relation in 1.5 keV WDM models result in reionization ending at comparable redshifts in all models. }  
\label{fig_qii} 
\end{figure}

\begin{figure*}
\center{\includegraphics[scale=0.92]{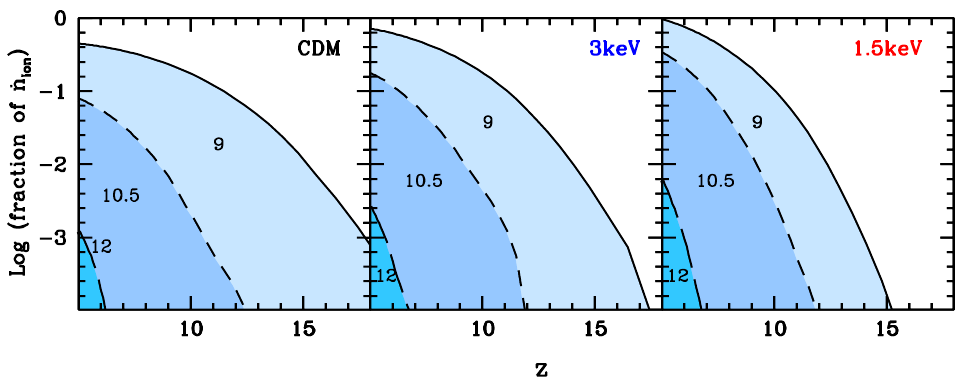}}
\caption{Fractional contribution from galaxies of halo masses $M_h \gsim 10^9, 10^{10.5}$ and $10^{12} \Msun$, as marked, for CDM (left panel), 3 keV (middle panel) and 1.5 keV WDM (right panel). As seen, galaxies with $M_h \gsim 10^9 \Msun$ provide about 40\% of the total ionizing budget, with the rest coming from lower mass halos in CDM. However, as a result of suppression of structure formation on small scales, these galaxies provide all of the ionizing photon budget in the 1.5 keV model. As expected, the contribution drops steeply with an increase in the halo mass such that $M_h \gsim 10^{12} \Msun$ galaxies contribute negligibly to the ionizing budget for all of the DM models considered. }
\label{fig_fracmh} 
\end{figure*}

\subsection{Reionization histories and sources in different DM models}
As discussed in Sec. \ref{sec_reio}, given the source galaxy population derived from the semi-analytic model, the reionization history, expressed through the global $z$-evolution of $Q_{II}$, is fixed once $f_{esc}(z)$ is determined. Given that CDM allows halo collapse on the smallest scales, it is the most affected by UVB feedback; the effect of UVB feedback naturally decreases with a decrease in the WDM particle mass. As shown in Fig. \ref{fig_qii}, reionization starts with the formation of the first galaxies as early as $z \simeq 18$ in CDM, corresponding to about 200 Myr after the Big Bang and is 50\% complete by $z \simeq 9$, initially driven by low mass $M_h \lsim M_{min}$ halos \citep[see also][]{bromm-yoshida2011}. Progressive UVB suppression of such halos with decreasing $z$ then results in $M_h \gsim M_{min}$ halos providing most of the \HI ionizing photons and reionization is 90\% complete by volume at $z \simeq 6$. 

Reionization is delayed by a small period about 60 Myr in $\mx \gsim 3\kev$ models compared to CDM as shown in the same figure. However, this minor delay is easily compensated by the steeper $f_{esc}$ behaviour (see Table \ref{table2}) such that reionization is 50\% and 90\% complete at, almost the same epochs as in CDM, at $z \simeq 8.5$ and $6$ respectively. With its largest suppression of low-mass halos, reionization only starts at $z \simeq 12$ in the 1.5 keV scenario, being delayed by about 150 Myr compared to CDM. This delay in the start of reionization is compensated by (i) the larger, and hence less feedback-limited, galaxies that produce stars and \HI ionizing photons at a much faster rate as compared to CDM and $\mx \gsim 3\kev$ models, and (ii) the steeper $f_{esc}$ evolution required to match to the observed optical depth and emissivity constraints. The two effects results in a faster reionization in the 1.5 keV model which is 90\% complete by $z \simeq 6.5$. Interestingly, reionization finishes by about $z \simeq 5.5$ in all these different scenarios, in accord with the results of \citet{yue2012} who have shown that although WDM can delay the start of reionization, the end is not necessarily delayed.

With regards to the main reionization sources, we now calculate the fractional contribution of galaxies to the total ionizing emissivity at $z \simeq 5$ as a function of their halo masses and magnitudes. Given the similarity between CDM and 5keV demonstrated above, for clarity, we only show results for CDM, 3 keV and 1.5 keV WDM in what follows. As a function of halo mass, we find that CDM galaxies with $M_h \gsim 10^9 \Msun$ provide only about $40\%$ of the total ionizing photons, with the dominant contribution coming from lower halo masses as shown in Fig. \ref{fig_fracmh}. Further, $M_h \gsim 10^{10.5} \Msun$ halos only provide $\sim 8\%$ of the total ionizing photons, dropping to less than 0.1 percent for $M_h \gsim 10^{12} \Msun$ halos. Given 3 keV WDM suppresses the formation of low-mass halos, $M_h \gsim 10^9 \Msun$ galaxies provide about 63\% of the total ionizing photons, with $M_h \gsim 10^{10.5} \,(10^{12}) \Msun$ halos contributing only 15\% (0.3\%) to the total ionizing budget. With the suppression of all halos below $\simeq 10^{9} \Msun$,  galaxies above this limit make up a 100\% of the ionizing photon budget in the 1.5 keV WDM model, with $M_h \gsim 10^{10.5} \Msun$ halos providing an appreciable 32\% of the total budget.

\begin{figure*}
\center{\includegraphics[scale=0.92]{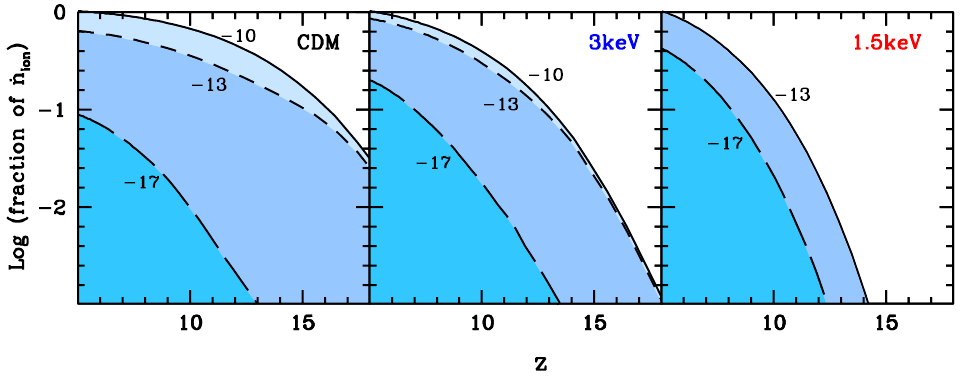}}
\caption{Fractional contribution from galaxies integrating down to UV magnitudes $M_{UV} = -10, -13$ and $-17$, as marked, for CDM (left panel), 3 keV (middle panel) and 1.5 keV WDM (right panel); given the lack of low-mass halos, there are no galaxies fainter than $M_{UV}=-13$ in the 1.5 keV WDM. As seen, while currently observed galaxies with $M_{UV} \lsim -17$ provide about 10\% of the total ionizing photon budget in CDM, this rises to about 40\% in the 1.5 keV scenario. Further, we find that while an integration limit of $M_{UV} = -13$ suffices for 1.5 keV WDM, a fainter limit of $M_{UV} = -10$ is required to catch all of the reionization photons for CDM and $\mx \gsim 3 \kev$ WDM. }
\label{fig_fracmuv} 
\end{figure*}

In terms of the observed UV magnitude, we find that an integration limit as faint as $M_{UV} =-10$ is required in order to catch the total ionizing photon budget for CDM and the 3 keV WDM models as shown in Fig. \ref{fig_fracmuv}; integrating down to the generally used limit of $M_{UV}=-13$ provides 63\% of the budget for CDM, rising to about 80\% for the 3 keV scenario, with the rest coming from fainter sources. Given the lack of low-mass halos, on the other hand, galaxies brighter than $M_{UV}=-13$ provide all of the ionizing budget for the 1.5 keV model. As expected from the discussion above, massive galaxies brighter than $M_{UV} = -17$ that reside in $M_h \gsim 10^{10.5} \Msun$ halos \citep[see Fig. 3;][]{dayal2014} provide only about 10\% of the total ionizing budget in CDM. This rises to about 20\% (40\%) for the 3 (1.5) keV models.

\begin{figure*}
\center{\includegraphics[scale=0.98]{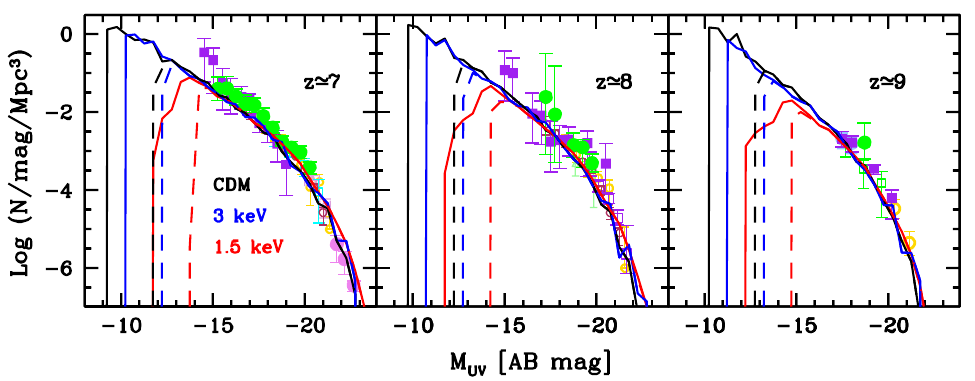}}
\caption{The evolving LBG UV LFs at $z \simeq 7$ (left panel), 8 (central panel) and 9 (right panel) for CDM, 3 keV and 1.5 keV WDM, as marked. In each panel, solid and dashed lines show results for SN feedback and the fiducial model (SN and UV feedback), respectively. As seen results from models with and without UV feedback are in agreement for all DM models, and with the observations, down to $M_{UV} \simeq -14$. In all panels points show observational results: (a) $z \simeq 7$: 
\citet[empty cyan squares]{oesch2010}, \citet[empty yellow circles]{bouwens2011b}, \citet[empty purple triangles]{castellano2010}, \citet[empty red triangles]{mclure2010}, \citet[empty maroon circles]{mclure2013}, \citet[filled purple circles]{bowler2014a}, \citet[filled green circles]{atek2015} and \citet[filled purple squares]{livermore2016}; (b) $z \simeq 8$: \citet[empty yellow circles]{bouwens2011b}, \citet[empty purple squares]{bradley2012}, \citet[empty maroon circles]{mclure2013}, \citet[filled green circles]{atek2015} and \citet[filled purple squares]{livermore2016} and (c) $z \simeq 9$: \citet[filled green circle]{zheng2012}, \citet[empty blue circles]{mclure2013}, \citet[empty green squares]{oesch2013}, \citet[filled purple squares]{mcleod2016} and \citet[empty yellow circles]{bouwens2016}.}
\label{fig_uvlf} 
\end{figure*}

To summarize, we find the bulk of the reionization photons come from galaxies with $M_h \lsim 10^9\Msun$ and $ -15 \lsim M_{UV} \lsim -10$ in CDM. The progressive suppression of low-mass halos with decreasing $\mx$ leads to a shift in the ``reionization" population to larger halo masses of $M_h \gsim 10^9\Msun$ and  $ -17 \lsim M_{UV} \lsim -13$ for $1.5\kev$ WDM. 

\section{Reionization impact on high-$z$ galaxy observables}
\label{impact}
We now quantify the impact of the UVB feedback on global observables such as the UV LF, the stellar mass density and the ionizing photon density, as well as observables measured for individual galaxies such as the stellar mass-halo mass relation. Given the small impact of changing $M_{min}$ between $10^{8.5}$ and $10^{9.5} \Msun$ on reionization noted in Sec. \ref{joint_cons}, we only show results for the fiducial case of $M_{min} = 10^9 \Msun$ in what follows. 

\subsection{Faint end of the UV luminosity functions}
\label{uvlf}

The evolving UV LF has now been measured over an unprecedented 10 (7) magnitudes at $z \simeq 7 \, (8)$ as shown in Fig. \ref{fig_uvlf}. At $z \simeq 7$ and 8, the faint end ($M_{UV}\gsim -16$) data has been collected using the lensed Hubble Frontier fields \citep[HFF;][]{atek2015, livermore2016} while the brightest galaxies have been detected using ground-based Ultra-Vista surveys \citep[e.g.][]{bowler2014a, bradley2013}; (unlensed) HST data has been used to obtain the $z \simeq 9$ UV LF as shown in the same figure. 

We start by discussing the UV LF considering only internal (SN) feedback; this effectively implies setting $M_{min}=0$ when modelling the feedback effects of the UVB. As shown, the UV LFs for CDM and 3 keV WDM are in agreement at least down to $M_{UV} \simeq -11$ at $z \simeq 7-9$; results from the 1.5 keV scenario are in agreement with these down to $M_{UV} \simeq -12$ at $z \simeq 7$ and start peeling-away at a slightly brighter magnitude corresponding to $M_{UV} \simeq -15$ at $z \simeq 9$ \citep[see also][]{dayal2014_wdm1}. However within error bars, all three UV LFs are in agreement with observations over the entire range probed so far. As shown in \citet{dayal2014_wdm1}, this overlap is the result of the absence of low-mass halos in WDM models being partly compensated by their younger, and hence brighter, stellar populations. 

Including the maximal effects of UV feedback, i.e. $M_{gas}=0$ and no star formation in all galaxies below the fiducial limit of $M_{min} = 10^9 \Msun$ at all $z$, naturally leads the cut-off shifting to brighter luminosities by about 2-3 AB magnitudes in all DM models considered: the $z \simeq 7$ LFs for CDM and 3 keV WDM cut-off at $M_{UV} \simeq -12$ while the 1.5 keV LF cuts-off at $M_{UV} \simeq -14$. It is interesting to see that probing down to $M_{UV} \simeq -14.5$, observations are beginning to enter the regime at which the 1.5 keV UV LF cuts-off, although all three DM models including maximal UV feedback are still in agreement with available data-sets. However, as shown in Fig. \ref{fig_qii}, roughly 85\% of the IGM is ionized by volume at $z \simeq 7$ in the 1.5 keV scenario- this implies 15\% of the volume would still be capable of hosting star formation in halos below $M_{min}$, pushing up the LF. Interestingly, the faint end slope remains unchanged for all DM models with $\alpha \simeq -2.04 \pm 0.74$ at $z \simeq 7$, increasing to $-2.10\pm0.62$ at $z \simeq 8$ \citep[see also][]{dayal2014}, which are in excellent agreement with the values of $\alpha \simeq -2.04 ^{+0.13}_{-0.17}$ observed at $z \simeq 7$ by \citet{atek2015} and of $\alpha \simeq -2.07 \pm 0.04 \, (-2.02\pm 0.04)$ observed at $z \simeq 7\, (8)$ by \citet{livermore2016}.

To summarise, CDM and WDM UV LFs overlap over the entire observed range and therefore can not be used to rule out WDM models even using the faintest galaxies with $M_{UV} \simeq -14.5$ detected so far, corresponding to halo masses $M_h \simeq 10^{9-9.5} \Msun$. Further, the exact magnitude at which the LF cuts-off depends both on the DM model as well as the feedback prescription implemented -  indeed, the 1.5 keV LF excluding UV feedback is essentially degenerate with that obtained in CDM including the effects of both SN and UV feedback.  

\begin{figure}
\center{\includegraphics[scale=0.47]{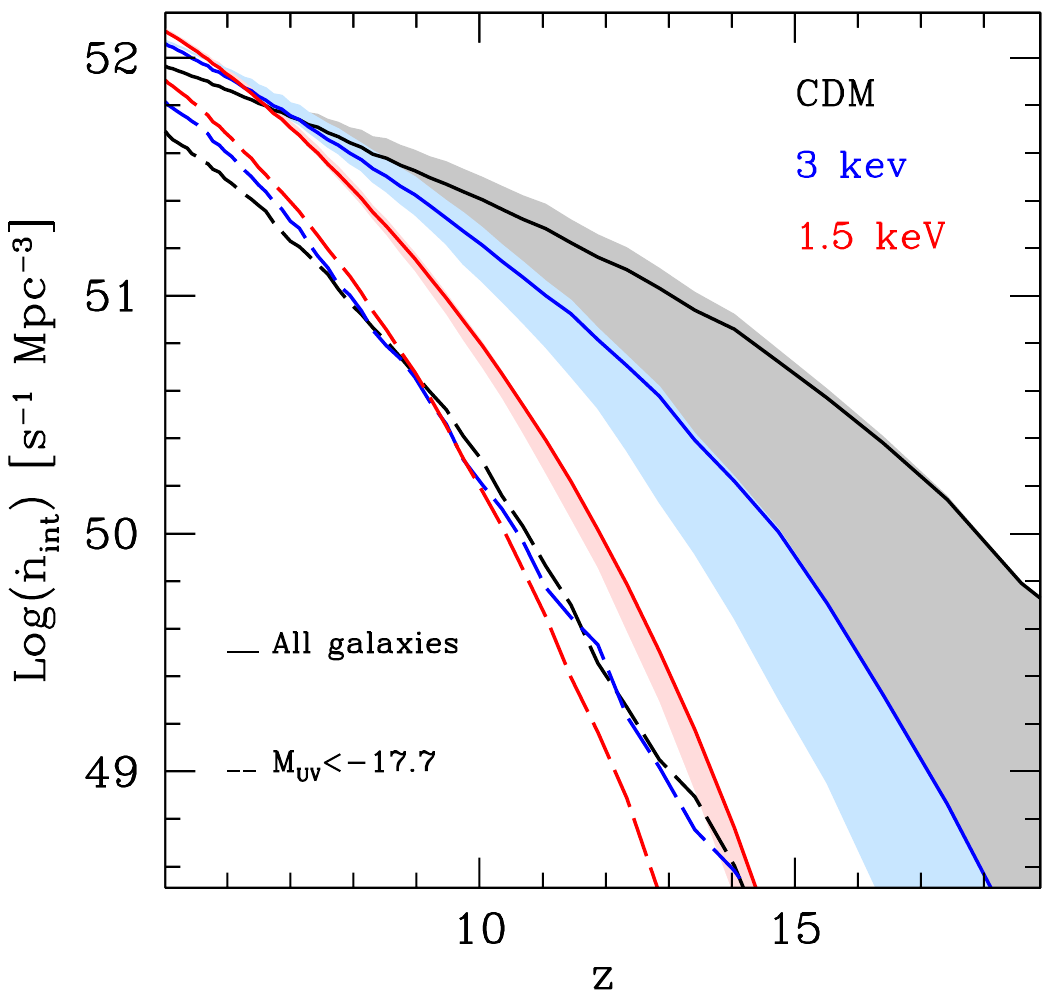}}
\caption{The redshift evolution of the intrinsic ionizing photon emissivity for CDM (black line), 3 keV (blue line) and 1.5 keV (red line) WDM. Solid lines show the fiducial results obtained using Eqn. \ref{eqn_nint} where a progressively larger fraction of low-mass ($M_h \lsim 10^9 \Msun$) halos are UV feedback suppressed, with dashed lines showing results for galaxies that have been detected with $M_{UV} \sim -17.7$ (dashed lines). The upper and lower limits of the shaded (black, blue and red) regions show the SMD range (for CDM, 3 and 1.5 keV WDM) obtained by ignoring and including the effects of UV feedback respectively. The $\dot n_{int}-z$ relations for different models are quantified in Table \ref{table3}. }
\label{fig_nint} 
\end{figure}

\begin{figure*}
\center{\includegraphics[scale=0.98]{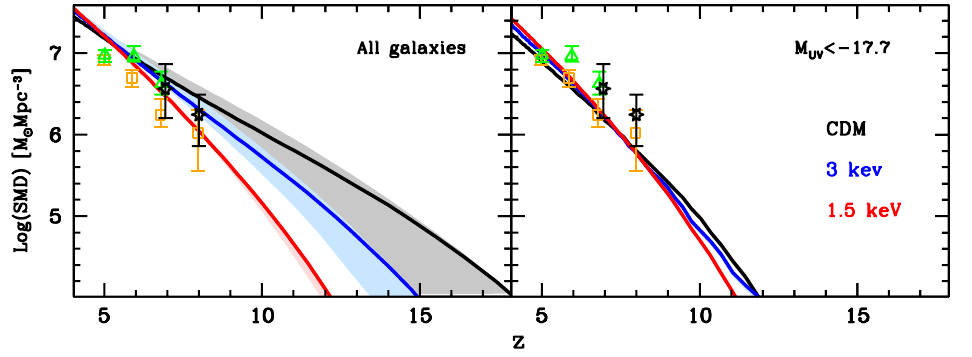}}
\caption{The SMD as a function of $z$ for CDM (black line), 3 keV (blue line) and 1.5 keV (red line) WDM for all galaxies (left panel) and galaxies that have been detected (right panel). The upper and lower limits of the shaded (black, blue and red) regions show the SMD range (for CDM, 3 and 1.5 keV WDM) obtained by ignoring and including the effects of UV feedback respectively. The solid lines show the fiducial results obtained using Eqn. \ref{eqn_smd} where a progressively larger fraction of low-mass ($M_h \lsim 10^9 \Msun$) halos are UV feedback suppressed. As seen from the right panel, the SMD from all three models are in accord for the massive galaxies ($\MUV \leq -17.7$) that have already been observed, shown by data points: \citet[empty triangles]{gonzalez2011}, \citet[empty squares]{stark2013} and \citet[empty stars]{labbe2010a,labbe2010b}.}
\label{fig_smd} 
\end{figure*}

\subsection{Ionizing photon density}
\label{nint}
The intrinsic ionizing photon density and its $z$-evolution is a key input for modelling reionization. We start by comparing $\dot n_{int}$ obtained from the different DM models considering three scenarios each of which includes SN feedback: (a) ignoring UV feedback i.e. $M_{min}=0$, (b) including maximal UV feedback, i.e. $M_{gas}=0$ in halos with $M_h \lsim M_{min}$ and (c) the more realistic case where we only suppress the gas mass of galaxies with $M_h \lsim M_{min}$ in ionized regions. We model the third case as 
\begin{equation}
\dot n_{int}(z) = \dot n_{int, ufb} Q_{II} + \dot n_{int,noufb} (1-Q_{II}). 
\label{eqn_nint}
\end{equation}
Here, $\dot n_{int,ufb}$ and $\dot n_{int,noufb}$ represent the ionizing photon production rates including and ignoring UV feedback, respectively. This formalism leads to a situation where a progressive fraction of $M_h \lsim M_{min}$ galaxies are photo-heating suppressed. UV feedback suppresses the largest amount of star formation and hence $\dot n_{int}$ in CDM - as shown in Fig. \ref{fig_nint}, including UV feedback decreases $\dot n_{int}$ by about  order of magnitude in the initial reionization stages at $z \gsim 15$. With a fewer number of low-mass halos, $\dot n_{int}$ is suppressed by a factor of 3 in 3 keV WDM at the same redshifts, while 1.5 keV WDM galaxies are essentially impervious to this effect. The $\dot n_{int}-z$ evolution for the fiducial model described in Eqn. \ref{eqn_nint} is tabulated in Table \ref{table3}. As noted in Sec. \ref{joint_cons}, the delay in structure formation, and hence the intrinsic production of ionizing photons, in the 1.5 keV WDM model can be compensated by a larger $f_{esc}$ at earlier epochs. Finally, we find that including/excluding UV feedback has no effect on $\dot n_{int}$ for the massive galaxies that have been observed with $M_{UV} \lsim -17.7$. Their large masses also result in a minimal dependence on the DM model, as shown in Fig. \ref{fig_nint} and tabulated in Table \ref{table3}.

\subsection{Stellar mass density (SMD)}
\label{smd}
Encoding information on the entire stellar mass assembled at any cosmic epoch, the SMD and its $z$-evolution is a key test of theoretical models. We discuss the SMD$-z$ relation considering the three key scenarios described in Sec. \ref{smd} above - no UV feedback, full baryonic suppression in galaxies with $M_h \lsim M_{min}$ and the realistic scenario wherein 
\begin{equation}
SMD(z) = SMD_{ufb} Q_{II} + SMD_{noufb} (1-Q_{II}). 
\label{eqn_smd}
\end{equation}

Here $SMD_{ufb}$ and $SMD_{noufb}$ represent the total SMD calculated using the maximal UV feedback scenario (case B above) and ignoring UV feedback (case A above), respectively. This formalism leads to a realistic situation where no galaxies are UV feedback suppressed in the initial reionization stages where $Q_{II}<<1$. As reionization progresses, a progressively larger fraction of galaxies below $M_{min}$ are UVB suppressed, culminating in all such galaxies being fully suppressed when $Q_{II} \rightarrow 1$. The results of these calculations for all galaxies at any $z$ are shown in the left panel in Fig. \ref{fig_smd} - as expected, including UV feedback has the largest effect on the SMD in the initial reionization stages for CDM, given it contains the largest number of low-mass halos below $M_{min}$. The effect of UV feedback reduces with decreasing redshift since such low-mass halos lose most/all of their gas mass after a single epoch of star formation, according to our formalism. The importance of UV feedback on the SMD naturally reduces with decreasing $\mx$ in WDM models given the dearth of low-mass halos. As seen from the same plot, while the SMD decreases by about an order of magnitude by including UV feedback for CDM, it only changes by about 0.5 (0.2) magnitudes for 3 (1.5) keV WDM. The $SMD-z$ relation for all the above mentioned cases is detailed in Table \ref{table3}. As clearly seen, the slope of the $SMD-z$ relation is extremely steep for the 1.5 keV model compared to CDM or 3 keV. As noted in \citep{dayal2014_wdm1} these relations are independent of the free-parameter (star formation efficiency threshold and fraction of SN energy coupling to winds) values used. Indeed, in the same work, the authors have shown that with its capability of pinning down the slope evolution of the SMD down to $M_{UV} \simeq -16.5$, the {\it James Webb Space Telescope} will be invaluable in differentiating between CDM and 1.5\kev\, WDM cosmologies.

\begin{figure*}
\center{\includegraphics[scale=0.98]{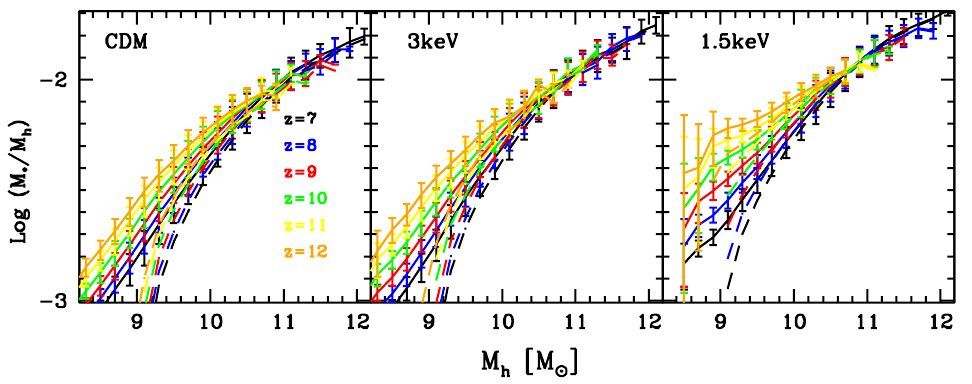}}
\caption{The $M_*/M_h$ relation as a function of halo mass for CDM (left panel), 3 keV (central panel) and 1.5 keV WDM (right panel) for the redshifts marked. In each panel, solid and dashed lines show results considering only SN feedback and the complete fiducial model (SN and UV feedback), respectively; we show $1-\sigma$ errors for the former case. As seen, the $M_*/M_h$ relation for low-mass ($M_h \lsim 10^{10}\Msun $) halos decreases in amplitude with decreasing redshift as a result of feedback progressively decreasing the gas mass, and hence star forming capabilities, of such halos; the relation converges for larger halo masses signifying the low importance of feedback on such massive systems. See Sec. \ref{smhm} for details.}
\label{fig_frac} 
\end{figure*}

\begin{figure*}
\center{\includegraphics[scale=0.98]{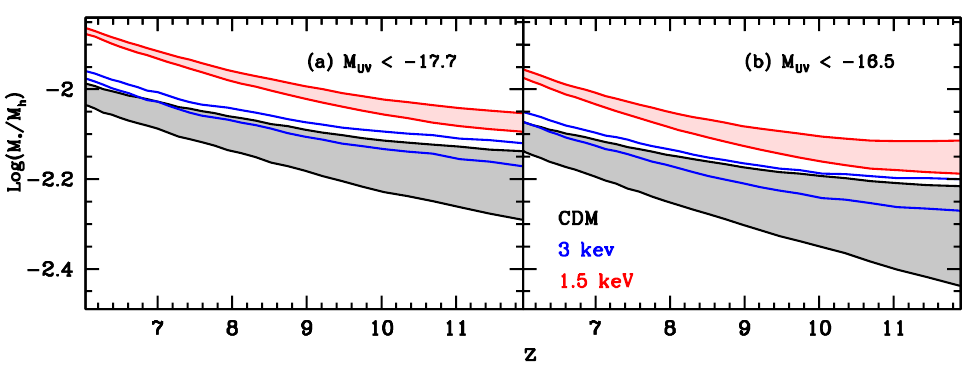}}
\caption{The $M_*/M_h$ relation as a function of $z$ for CDM (black lines), 3 keV WDM (blue lines) and 1.5 keV WDM (red lines) integrating down to a UV magnitude corresponding to $-17.7$ (observed galaxies) and $-16.5$ in the eft and right panels, respectively. In each panel, the top and bottom lines show the relation ignoring and including UV feedback, for each DM model. As seen, the steeper slope of the $M_*/M_h -z$ relation could be used to obtain constraints on the WDM particle mass. }
\label{fig_msbymh} 
\end{figure*}

Finally, we check that the SMD from all 3 DM models are in accord with observed data, as shown in the right panel of the same figure. The observed SMDs have been inferred by integrating down to a limit of $M_{UV} \lsim -17.7$. As expected these galaxies, corresponding to $M_h \lsim 10^{10.5} \Msun$, are large enough to be impervious to the effects of UV feedback, as a result of which $SMD_{ufb} \simeq SMD_{noufb}$. The theoretical $SMD-z$ relations for observed galaxies are also tabulated in Table \ref{table3}.

\subsection{Stellar mass-halo mass relations}
\label{smhm}
We now discuss the relation between the stellar and halo mass for high-$z$ galaxies for the three DM models considered. As expected, the value of $M_*/M_h$ increases with increasing halo mass for all models for $z \simeq 7-12$, signifying their larger star formation efficiencies. Starting by ignoring UVB feedback, in CDM, $z \simeq 7$ $M_h \simeq 10^9 \Msun$ galaxies have $M_*/M_h \simeq 0.2\%$ - this rises by a factor of about 7 to $1.4\%$ for $10^{11.5} \Msun$ halos, with the slope flattening off above this value. While the $M_*/M_h$ value is independent of $z$ for $M_h \gsim 10^{10.5}$ galaxies, corresponding to $M_{UV} \lsim -17$, $M_*/M_h$ decreases with decreasing $z$ for lower mass halos by about 0.3 dex from $z \simeq 12$ to 7. This is a result of SN-driven winds progressively decreasing the gas mass, and hence star-formation capabilities, in low-mass halos through time. While a similar behaviour is observed for $M_h \gsim 10^{10.5}$ galaxies in 1.5 \kev\, WDM, lower mass galaxies exhibit higher $M_*/M_h$ values at earlier cosmic epochs as a result of the larger (initial) potential wells - $M_*/M_h$ rises from about 0.2\% at $z \simeq 7$ to about 0.6\% at $z \simeq 12$ for $M_h \simeq 10^9 \Msun$ galaxies in this model, compared to the marginal rise from 0.2\% to 0.3\% seen in CDM at the same redshifts. 

Including the maximal UV feedback scenario (i.e. $M_{gas}=0$ for $M_h \lsim 10^9 \Msun$ halos) naturally results in a decrease in the $M_*/M_h$ ratio for low mass halos since their (gas devoid) progenitors are unable to form stars and build up their stellar content. This results in the $M_*/M_h$ values of low-mass ($M_h \lsim 10^{9.5}\Msun$) galaxies decreasing by about 0.3 dex in CDM and about 0.15 dex for 1.5 keV WDM at $z \simeq 12$. We find that the effect of UV feedback loses its impact with decreasing $z$ as a result of low-mass progenitors losing gas due to SN feedback. While WDM galaxies show the same trends of $M_*/M_h$ with halo mass and $z$, there are quantitative differences: firstly, the $M_*/M_h$ trend does not show any flattening at large masses which is a result of galaxies starting from larger, and hence less feedback-limited, progenitors in such models compared to CDM. Secondly, at a given halo mass, while the CDM and 3 keV $M_*/M_h$ relations are quite similar, the amplitude of this relation is higher by about 0.2-0.4 dex in the 1.5 keV scenario which is again driven by the fact that starting from larger progenitors, galaxies of a given halo mass are less feedback limited and hence able to assemble larger stellar masses. We end by noting that the ``true" stellar mass value should lie between these limits of zero and maximal UV feedback; while most low-mass galaxies would be unaffected by UV feedback in the initial reionization stages, the results would move closer to the maximal UV feedback scenario in the end reionization stages. 

We then study the $z$ evolution of the $M_*/M_h$ relation in different DM cosmologies, shown in Fig. \ref{fig_msbymh}. As seen from this plot, CDM shows the lowest $M_*/M_h$ value at any $z$ - this is the result of stronger feedback decreasing the total stellar mass of low-mass halos. As expected, the  $M_*/M_h$ value increases with decreasing $\mx$ at any redshift. Indeed, integrating down to a limit of $M_{UV} \simeq -17.7$, CDM galaxies show a $M
_*/M_h-z$ slope proportional to $0.003z^2 \, (0.002z^2)$ including (excluding) UV feedback. The corresponding, steeper, 1.5 keV WDM slopes have a value of about $0.004 z^2$, given the minimal effect of UV feedback on galaxies in this model. As expected, the slopes steepen in all cases integrating down to a fainter magnitude limit of $M_{UV} = -16.5$. In this case, the CDM slopes correspond to $0.003 z^2 \, (0.0015z^2)$ with (without) UV feedback which steepen to $0.005z^2$ for the 1.5 keV model as tabulated in Table \ref{table3}. Current estimates on the stellar mass have an error of about 0.3 dex, with abundance matching techniques, to infer halo masses, resulting in errors of the order of 0.5 dex. While current stellar mass-halo mass observations are probably unable to distinguish between these different slopes, more precise estimates, of both stellar and halo masses, in the future might offer a useful tool to distinguish between CDM and 1.5 keV WDM. 

\section{Conclusions and discussion}
Although the Cold dark matter (CDM) paradigm has been enormously successful in explaining the large scale structure of the universe, it shows a number of small scale problems that can be alleviated by invoking (\kev) Warm Dark Matter (WDM) particles that erase small scale power, suppressing the formation of low-mass structures. We use a semi-analytic model traces the DM and baryonic assembly of high-$z$ ($z\simeq 5-20$) galaxies in 4 DM cosmologies - CDM and WDM with $\mx = 1.5,3$ and $5\kev$ - including the key baryonic processes of star formation, SN feedback and the merger/accretion/ejection driven evolution of gas and stellar mass. This model uses only two {\it $z-$ and mass-independent free parameters} with fiducial values of 3.8\% for the maximum star formation efficiency and 10\% of the SN energy going into unbinding gas \citep{dayal2014, dayal2014_wdm1}. Modelling reionization requires an additional free parameter which is the fraction of ionzing photons that can escape out of the galactic environment, $f_{esc}$. In this work, we also include the effects of external feedback from a UVB suppressing the baryonic content of galaxies below a threshold mass, for which we use a fiducial value of $M_h \simeq 10^9\Msun$. The model is constrained using key observables for high-$z$ galaxies (UV LF, SMD, M/L ratios), the latest constraints on the cosmological electron-scattering optical depth ($\tau_{es}$) from {\it Planck} and the measured ionizing photon emissivity ($\dot n_{ion}$) to understand: (i) the reionization history and sources , and (ii) the effects of UV feedback on galaxy assembly in different DM cosmologies. 
Our main findings can be summarized as follows:

\begin{itemize}
 
\item Simultaneously matching $\tau_{es}$ and $\dot n_{ion}$ requires a $z-$dependent $f_{esc} \propto (1+z)^\alpha$. Although 1.5 keV WDM requires an exceptionally steep $z-$dependence, compared to the other models, the latest $\tau_{es}$ estimates \citep{planck2015,planck2016} allow the DM mass to range from ~Gev for CDM to 1.5 keV WDM. 

\item Although the start of the reionization process is delayed in light WDM models, a faster galaxy assembly and a steeper redshift dependence of $f_{esc}$ result in reionization ending at almost the same epoch $z \simeq 5.5-6$ in all DM models \citep[see also][]{yue2012}. In terms of reionization sources, the bulk of the reionization photons come from galaxies with $M_h \lsim 10^9\Msun$ and $ -10 \lsim M_{UV} \lsim -15$ in CDM. The progressive suppression of low-mass halos in light WDM models results in a shift in the reionization population to larger halo masses of $M_h \gsim 10^9\Msun$ and $ -13 \lsim M_{UV} \lsim -17$ for $\mx = 1.5\kev$ WDM. 

\item We quantify the effects of UV feedback on observables including the evolving UV LF, SMD, $\dot n_{int}$ and the $M_*/M_h$ ratio and show that its effects are the most pronounced for CDM (with the largest number of low-mass halos) and decrease with a decreasing WDM particle mass. We find that even in the maximal UV feedback scenario (where $M_h \lsim 10^9 \Msun$ galaxies have no gas content), all DM models (CDM, 3 and 1.5 keV WDM) are in agreement with each other down to the current observational limits of $M_{UV} \simeq -14.5$.

\item Interestingly, the UV LFs obtained from the different DM models considered are degenerate with the physics implemented - for example, the 1.5 \kev LF excluding UV feedback is indistinguishable from that obtained in CDM including the effects of both SN and UV feedback. We find that the SMD and $\dot n_{ion}$ values are indistinguishable amongst the different DM models for observed galaxies, with the differences being the most pronounced for the smallest halos.

\item We find that, in addition to the slope of the $SMD-z$ relation which will be measured with the forthcoming {\it JWST} \citep{dayal2014_wdm1}, another global estimate is the redshift evolution of the $M_*/M_h$ relation that can potentially be used to put constraints on the DM particle mass.

\end{itemize}

We end with a few caveats. Firstly, we have assumed reionization to be solely driven by the stars within galaxies. In recent times, it has been proposed that a population of relatively faint quasars at high-$z$ can complete reionization without any significant contribution from galaxies \citep{madau2015, khaire2016, mitra2015, daloisio2016}. It would be interesting to explore the contribution of these quasars in the framework of the WDM models and see if these objects can relax the rapidly evolving $f_{esc}$ required from 1.5 \kev\, WDM models. Secondly, we have ignored SN radiative losses that could significantly reduce the total energy available to drive winds. However, a decrease in this total energy could be countered by scaling up the fraction of the total energy we put into driving winds from the fiducial value of 10\% used in this work. Thirdly, \citet{pawlik2009} have shown that UVB photo-heating reduces the clumping factor of the IGM since the additional pressure support from reionization smoothes out small-scales density fluctuations. While we have used their results for an over-density of 100, we have confirmed that our results do not change using threshold values of 50, or 200. Fourthly, while we assume that all halos with mass $M_h \lsim 10^9\Msun$ are feedback suppressed, we find our results are equally consistent with observations ($\tau_{es}$ and $\dot n_{ion}$) using values ranging between $10^{8.5}-10^{9.5}\Msun$. Finally, we have made the simplifying assumption of using a mass-independent $f_{esc}$ at all $z$. This is partly motivated by the uncertainty regarding the mass-dependence of $f_{esc}$: while some authors find $\fesc$ to decrease with an increase in the halo mass \citep{razoumov2010,yajima2011,ferrara2013}, other works have found the opposite trend \citep{gnedin2008,wise2009}. Significant progress is expected to be made by comparing our model predictions with further LBG data expected from the Frontier Fields, and from forthcoming observatories such as the {\it JWST}. Furthermore, if the claim of $3.5$~keV X-ray line holds to be true, it will usher in an exciting era for WDM dominated cosmology.


\begin{table*}
\begin{center}
\caption{Relations between different quantities for the 3 DM models considered}
\begin{tabular}{|c|c|c|c|c|}
\hline
DM Model & SN feedback & UVB feedback & UV cut & $\dot n_{int}-M_*$ relation at $z \simeq 7$\\
\hline
CDM & yes & no &  all galaxies & $\log (\dot n_{int}) = \log(M_*) + 45$\\
3 keV WDM & yes & no &  all galaxies & $\log (\dot n_{int}) = 0.98 \log(M_*) + 45$\\
1.5 keV WDM & yes & no &  all galaxies & $\log (\dot n_{int}) = 0.88 \log(M_*) + 46$\\
CDM & yes & yes &  all galaxies & $\log (\dot n_{int}) = 0.95 \log(M_*) + 46$\\
3 keV WDM & yes & yes &  all galaxies & $\log (\dot n_{int}) = 0.91 \log(M_*) + 46$\\
1.5 keV WDM & yes & yes &  all galaxies & $\log (\dot n_{int}) = 0.85 \log(M_*) + 47$\\

\hline

DM Model & SN feedback & UVB feedback & UV cut & $SMD-z$ relation\\
\hline
CDM & yes & no &  all galaxies & $\log (SMD) = -0.0038 z^2 -0.16 z+8.2$\\
CDM & yes & yes &  all galaxies & $\log (SMD) = -0.0086 z^2 -0.16 z+8.2$\\
CDM & yes & yes &  all galaxies (Eqn. \ref{eqn_smd}) & $\log (SMD) = -0.0015 z^2 -0.21 z+8.2$\\

3 keV & yes & no &  all galaxies & $\log (SMD) = -0.011 z^2 -0.18 z+8.4$\\
3 keV & yes & yes &  all galaxies & $\log (SMD) = -0.006 z^2 -0.2 z+8.4$\\
3 keV & yes & yes &  all galaxies (Eqn. \ref{eqn_smd}) & $\log (SMD) = -0.0032 z^2 -0.25 z+8.6$\\

1.5 keV & yes & no &  all galaxies & $\log (SMD) = -0.02 z^2 -0.13 z+8.4$\\
1.5 keV & yes & yes &  all galaxies & $\log (SMD) = -0.02 z^2 -0.14 z+8.4$\\
1.5 keV & yes & yes &  all galaxies (Eqn. \ref{eqn_smd}) & $\log (SMD) = -0.017 z^2 -0.16 z+8.4$\\

CDM & yes & yes &  $M_{UV} \lsim -17.7$ & $\log (SMD) = -0.011 z^2 -0.23 z+8.4$\\
3 keV & yes & yes &  $M_{UV} \lsim -17.7$ & $\log (SMD) = -0.014 z^2 -0.21 z+8.4$\\
1.5 keV keV & yes & yes &  $M_{UV} \lsim -17.7$ & $\log (SMD) = -0.026 z^2 -0.085 z+8.1$\\
\hline

DM Model & SN feedback & UVB feedback & UV cut & $M_*/M_h-z$ relation\\
\hline
CDM & yes & no &  $M_{UV} \lsim -16.5$ & $\log (M_*/M_h) = -0.0031z^2 - 0.018 z-1.7$\\
CDM & yes & no &  $M_{UV} \lsim -17.7$ & $\log (M_*/M_h) = -0.0031z^2 - 0.08 z-1.6$\\
1.5 keV & yes & no & $M_{UV} \lsim -16.5$ & $\log (M_*/M_h) = -0.0054z^2 - 0.012 z-1.4$\\
1.5 keV & yes & no & $M_{UV} \lsim -17.7$ & $\log (M_*/M_h) = -0.0041z^2 - 0.01 z-1.4$\\

CDM & yes & yes &  $M_{UV} \lsim -16.5$ & $\log (M_*/M_h) = -0.0015z^2 - 0.07 z-1.7$\\
CDM & yes & yes &  $M_{UV} \lsim -17.7$ & $\log (M_*/M_h) = -0.0021z^2 - 0.08 z-1.6$\\
1.5 keV & yes & yes &  $M_{UV} \lsim -16.5$ & $\log (M_*/M_h) = -0.0051z^2 - 0.013 z-1.4$\\
1.5 keV & yes & yes &  $M_{UV} \lsim -17.7$ & $\log (M_*/M_h) = -0.004z^2 - 0.011 z-1.4$\\
\hline

\end{tabular}
\label{table3}
\end{center}
\end{table*}

\section*{Acknowledgments} 
PD acknowledges support from the European Commission's CO-FUND Rosalind Franklin program. VB acknowledges support from NSF grant AST-1413501. FP acknowledges the SAO Chandra grant AR6-17017B and NASA-ADAP grant MA160009. PD and TRC thank the Sexten Centre for Astrophysics for their hospitality, where a part of this work was carried out. The authors thank A. Taylor, M. Viel for helpful discussions and A. Mazumdar for help with understanding the particle physics aspect of WDM.


\bibliographystyle{apj}
\bibliography{reio}

\label{lastpage} 
\end{document}